# Controls at the Fermilab PIP-II Superconducting Linac *

Dennis J. Nicklaus, Pierrick Hanlet, Charles King, , Daniel McArthur, Richard Neswold, Fermilab, Batavia, IL, USA

*Abstract*

PIP-II is an 800 MeV superconducting RF linear accelerator under development at Fermilab. As the new first stage in our accelerator chain, it will deliver high-power beam to multiple experiments simultaneously and thus drive Fermilab's particle physics program for years to come. In a pivot for Fermilab, controls for PIP-II are based on EPICS instead of ACNET, the legacy control system for accelerators at the lab. This paper discusses the status of the EPICS controls work for PIP-II. We describe the EPICS tools selected for our system and the experience of operators new to EPICS. We introduce our continuous integration / continuous development environment. We also describe some efforts at cooperation between EPICS and ACNET, or efforts to move towards a unified interface that can apply to both control systems.

## PIP-II INTRODUCTION

PIP-II is a new superconducting radio frequency linear accelerator under construction at Fermilab. It will produce a pulsed 800MeV high intensity pulsed H- beam to supply the needs of Fermilab Long Baseline Neutrino Facility experiments and more. The Linac is followed by a beam transport line to deliver the beam to the Booster. Key components of the Linac include:

- Ion Source
- Low Energy Beam Transport
- RF Quadrupole to accelerate to 2.1 MeV
- Medium Energy Beam Transport with a chopper system to form the required bunch structure
- One 162.5 MHz Half Wave Resonator Cryomodule
- Two 325 Single Spoke Resonator Cryomodule (SSR1)
- Seven 325 Single Spoke Resonator Cryomodule (SSR2)
- Eleven Low Beta 650 MHZ cryomodules
- Four High Beta 650 MHz cryomodules

Of course, there is a considerable amount of instrumentation and beam controlling magnets in between the variety of cryomodules, as well as infrastructure to supply the cryogenics required to operate the superconducting cavities at 2K.

## CONTROLS OVERVIEW

EPICS has been selected as the central control system for PIP-II. However, the nature of the EPICS toolkit means that there are still many choices to be made for key elements of the system. Furthermore, while PIP-II is a new machine without legacy controls to hinder the adoption of EPICS, it will feed into the Fermilab Booster accelerator which is controlled by Fermilab's legacy ACNET[1] control system. While it would be technically possible for the two control systems to simply co-exist in our control rooms, operation efficiency is going to require that there is a fair amount of cooperation and unity between PIP-II's EPICS and the rest of the complex's ACNET.

*Modernization Cooperation*

Fermilab is also beginning a major modernization effort for the legacy control system, known by the project acronym ACORN (Accelerator Controls Operations Research Network)[2]. ACORN is looking at improving all facets of the control system, from the software frameworks to fieldbus hardware to user interfaces. The later schedule of the modernization effort means that PIP-II has a policy of not depending on any result or software that ACORN will produce. It also adds a slight mental burden to PIP-II controls development because we don't want to make any decisions that are drastically conflicting with likely ACORN directions. However, PIP-II is able to synergistically work with ACORN in some respects, taking advantage of research done by ACORN. One instance of this was when both projects wanted to evaluate software technologies for web-based controls applications. PIP-II began prototyping after ACORN selected a combination of Dart language and Flutter for application user interface building.

## EPICS COMPONENTS

*IOCs*

PIP-II will have many IOCs (Input Output Controllers) built off EPICS base as the data collection end of the control system. These IOCs will be built with EPICS version 7 to enable the PV Access protocol. In a major change for Fermilab, we don't foresee any VME-based embedded computers running a real time operating system. While it is an architecture that our hardware and software engineers are very comfortable with, the price/performance of these embedded computer cards has become extremely poor in recent years, so we no longer see a use case for these systems. We expect that most of our IOCs will be "soft" IOCs, server based and communicating with field hardware over ethernet. Many of the subsystems will be FPGA (including "system on a module" (SOM) style subsystems). We may eventually embed an IOC into some of these subsystems, but at this moment, most of our plans call for the IOC to be on a remote server, using some other (non-PV Access) Ethernet protocol to communicate with the FPGA systems.

For higher bandwidth instrumentation systems, e.g., BPMs (Beam Position Monitors), BLMs (Beam Loss Monitors), current monitors, etc. we are planning to use MicroTCA bus-based systems. But here again, since the MicroTCA bus and the data acquisition cards can support ethernet protocols, we do not plan to utilize an in-crate CPU card. Rather, we will have an isolated high speed ethernet linking the data acquisition cards to a more capable server computer. This allows us to have appropriately powerful and scalable server available at common commodity computing prices. It is much easier to upgrade the commodity computing servers to stay current with operating systems or hardware capabilities. We may one day find a case where an in-crate CPU is desired, but there are no current designs including one under consideration.

### Standard EPICS Services

One of the primary reasons for selecting EPICS for PIP-II was the suite of software services developed and supported by other laboratories. We have adopted several of these standard EPICS products for use at PIP-II. These include the Archiver Appliance, Channel Finder, and the Alarm Server (the Kafka based instance, integrated into Phoebus), Save & Restore, and, notably, the Phoebus version of Control System Studio. While some of these, such as the Archiver Appliance have been implemented without local modification, we have made our own local additions or changes to some of the software to suit our environment.

Operations of our initial test stand for PIP-II, called PIP2IT, relied heavily on Phoebus for graphical user displays and alarm displays. We've made an extension to Phoebus by adding a "plug in" that communicates with ACNET devices over the ACNET protocol. This allows us to access ACNET devices on a Phoebus screen by specifying them using an "acnet" prefix.

We expect Channel Finder to be useful in its basic form, but we also foresee extensions to its functionality by using its programming interfaces to connect to programs like our ACNET database or ACNET Data Pool Manager to supplement the information they use or to validate data requests or to check the syntax of PV names.

## ACNET – EPICS COOPERATION

As mentioned above, EPICS and ACNET will need to work together once the PIP-II beam is injected into the Booster accelerator. In order to present operators and engineers with a more straightforward control system, we are taking some steps to modify both ACNET and some EPICS software.

### Data Pool Manager

Central to this interactio is the existing control system's Data Pool Manager (DPM). The DPM manages control system requests, diverting them to the appropriate "front end" node (ACNET's equivalent of an IOC) for fulfilment. Part of DPM's functionality is to pool requests to the same front end for efficiency. We modified DPM so that it can also communicate via PV Access to EPICS IOCs. So, it must determine whether a requested device or PV belongs to ACNET or EPICS, then route the request accordingly.

### Authentication

A key design element of the control system is its support for authentication and authorization. Here, authentication is how the user proves he is who he claims to be and how those credentials are presented to the control system software. Authorization determines which activities (e.g., settings, accessing certain parts of the machine) an authenticated user is allowed. Fermilab has a culture of allowing operators, engineers, scientists, or technicians to run control system consoles and applications from their computers anywhere on site, in addition to the support for the accelerator complex's Main Control Room and outlying control rooms.

Our design calls for both EPICS and ACNET traffic to be routed through the DPM. Thus, DPM will be tasked with verifying the user's credentials. This does make DPM a potential bottleneck, but DPM is designed to run on multiple hosts simultaneously and to automatically load balance. If it becomes too busy, we can add additional DPM nodes.

We are currently working with Keycloak software to provide the authentication layer. This can work with Fermilab's existing Single Sign On (SSO) infrastructure to give users a familiar interface and reduce the sign-on burden and complexity.

We have not yet begun the addition of this authentication method to Phoebus but are gaining experience with it using the Dart-Flutter web application environment mentioned above.

### Alarms

We are working to integrate the ACNET and EPICS alarm services so that the operators only need to monitor one system. The initial path for this seems to be to pass along any ACNET alarms to the EPICS Alarm Server for handling and display. This entails more than simply translating the ACNET message to an EPICS format. For instance, there is a substantial database of alarm configuration in the ACNET database that will need to be used to inform the EPICS configurations.

### User Interfaces

As mentioned previously we have already added an ACNET plugin to Phoebus so that Phoebus can access ACNET data. We're also using Dart and Flutter to build prototype web-based applications that can access either data source through our DPM.

## TIMING

We are also implementing a customized timing system for PIP-II to coordinate the operation of the Linac and the rest of the existing accelerator complex by distributing the required clocks, machine resets, triggers, and system state information. It is being implemented as a two-part system – a more global timing system that supplies the whole Fermilab accelerator complex (ACLK) and an RF synchronized clock system unique to the PIP-II Linac (LCLK). The two systems will be implemented with comparable hardware. LCLK will have a 650 MHz carrier frequency distributed across the machine by fibre and be synchronous to the PIP-II Linac's LLRF. ACLK and LCLK use the same "modified Manchester" encoding scheme that the existing accelerator's clock system, TCLK, uses. ACLK and LCLK packets will consist of a 16-bit Event field and a 32-bit Data field. The data field for each event is not yet defined. By default, it is expected to be an "event stamp", i.e., a counter that increments with each occurrence of the event, but the data field can be redefined to most anything that will fit into 32 bits, such as a timestamp, machine data, etc.

## CONCLUSION

We are primarily implementing an EPICS based control system for PIP-II, Fermilab's powerful new initial accelerator stage. However, we are also implementing several customizations and extensions to adapt EPICS to the unique needs of our site.

## ACKNOWLEDGEMENT

This manuscript has been authored by Fermi Research Alliance, LLC under Contract No. DE-AC02-07CH11359 with the U.S. Department of Energy, Office of Science, Office of High Energy Physics.

Fermilab paper # FERMILAB-CONF-23-540-AD.